# Computing on Masked Data: a High Performance Method for Improving Big Data Veracity

Jeremy Kepner, Vijay Gadepally, Pete Michaleas, Nabil Schear, Mayank Varia, Arkady Yerukhimovich, Robert K. Cunningham

MIT Lincoln Laboratory, Lexington, MA, U.S.A.

*Abstract—* The growing gap between data and users calls for innovative tools that address the challenges faced by big data volume, velocity and variety. Along with these standard three V's of big data, an emerging fourth "V" is veracity, which addresses the confidentiality, integrity, and availability of the data. Traditional cryptographic techniques that ensure the veracity of data can have overheads that are too large to apply to big data. This work introduces a new technique called Computing on Masked Data (CMD), which improves data veracity by allowing computations to be performed directly on masked data and ensuring that only authorized recipients can unmask the data. Using the sparse linear algebra of associative arrays, CMD can be performed with significantly less overhead than other approaches while still supporting a wide range of linear algebraic operations on the masked data. Databases with strong support of sparse operations, such as SciDB or Apache Accumulo, are ideally suited to this technique. Examples are shown for the application of CMD to a complex DNA matching algorithm and to database operations over social media data.

## I.  Introduction

Big data and big data solutions are commonly characterized by the three V's: volume, velocity, and variety [Laney 2001]. The 17 member agencies of the US intelligence community face many similar issues in transforming their big data into actionable solutions for their users. Led by the National Security Agency, a Common Big Data Architecture (CBDA) was developed based on the Google Big Table design [Chang 2008] to help address these issues and is now in wide use. The centerpiece of the CBDA is the Apache Accumulo database (accumulo.apache.org) and its D4M (aka NuWave) schema [Kepner 2013a]. The CBDA connects users with data using a range of technologies: file systems, ingest processes, databases, data analytics, web services, scheduling, and elastic computing (Figure 1). Increasingly, big data solutions must address the confidentiality, integrity and availability of their data leading to the designation of a fourth V: veracity.

There are many veracity challenges in a big data system (see Figure 1a): external denial of service, credential stealing, cross virtual machine (VM) side channels, VM hypervisor privilege escalation, remote code injection, data integrity attacks, data loss/exfiltration, insider threats, internal network resource attacks, and supply chain attacks [Evans 2013]. These attacks threaten the availability, confidentiality, and integrity of both the original data and the analytic results.

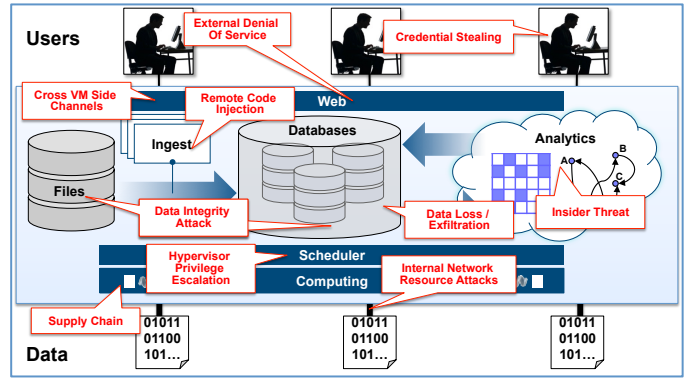
(a) big data veracity: challenges

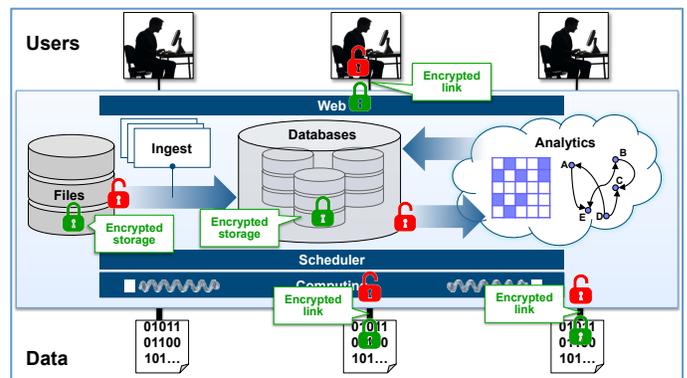
(b) big data veracity: current approaches

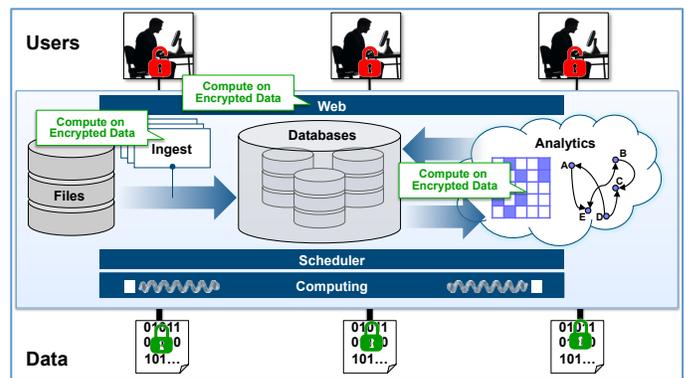
(c) big data veracity: vision

Figure 1. The Common Big Data Architecture connects diverse users with diverse data. Shown are big data veracity challenges (a), current approaches (b), and longer term vision (c).

These attacks threaten to degrade the availability of the big data system, compromise the confidentiality of the data and analytics used, and violate the integrity of both the original data and the analytic results. There are several approaches to mitigating these veracity challenges. Data centric veracity protections are particularly useful for preserving the confidentiality of the data. Typical defenses of this type include (see Figure 1b): encrypting the links between users and the big data system, encrypting the links between the data sources and the big data system, encrypting the data in the file system, and encrypting data in the database when it is at rest. These approaches are all significant steps forward in improving the veracity of a big data system. However, all of these approaches require that the data be decrypted for it to be used inside the big data system, which requires that the keys to the data be available to the big data system thus exposing the data to any attacker able to breach the boundaries of the system.

One vision (Figure 1c) for a big data system is to have data sources encrypt data prior to transmitting it to the system, have the big data system operate on the data in encrypted form, and only allow authorized users the keys to decrypt the answer for their specific result. Such a system makes the underlying big data technologies oblivious to the details of the data and would go a long way towards mitigating the big data veracity challenges described above. As a result, the data and processing can be outsourced to an untrusted cloud while preserving the confidentiality of the data and results.

Our Computing on Masked Data (CMD) system is a first step in this direction allowing for basic computations on encrypted data to enable a rich class of analytics. CMD combines efficient cryptographic encryption methods with an associative array representation of big data to enable a low computation cost approach to both computation and query while revealing only a small amount of information about the underlying data. The overhead of CMD is sufficiently low (~2x) to make it feasible for big data systems. Currently, all big data systems must operate on their data in the clear. CMD raises the bar by enabling some important computations on encrypted data while not dramatically increasing the computing resources required to perform those operations.

The outline of the rest of this paper is as follows. Section II gives a brief introduction to the relevant cryptographic tools. Section III describes the mathematics of associative arrays that underpin CMD and the D4M schema that is used to transform big data into associative arrays. Section IV describes the CMD system and how it can be integrated with the standard big data schemas. Section V presents the results of using CMD for analyses on bioinformatics and social media data. Section VI presents our conclusions and plans for future work.

## II. RELEVANT CRYPTOGRAPHIC TOOLS

There are several cryptographic tools that one could use to build a system like CMD. First, fully homomorphic encryption (FHE) allows for arbitrary analytic computations to be performed on encrypted data without decrypting it and while preserving its semantic security (i.e., no information about the data is leaked other than its length). FHE has been an active topic of research since its discovery [Gentry 2009]. Nevertheless, the best currently available schemes [Perl 2011, Halevi 2014] have an overhead of $10^5$ or more, making them too slow for use in practical big data systems.

If one is willing to allow a small amount of information about the encrypted data to be revealed, a much more efficient alternative to using FHE is to design protocols that leverage more traditional cryptographic techniques to carry out queries on encrypted data. One example of such a protocol is CryptDB [Popa 2011], which constructs a practical database system capable of handling most types of SQL queries on encrypted data. It uses deterministic encryption (DET), which always encrypts the same data to the same ciphertext, to enable equality queries; order-preserving encryption (OPE), which encrypts data in a way that preserves the original order of the data, to enable range queries; and additively homomorphic encryption (HOM+), which enables summing values directly on encrypted data, to perform basic analytics. Several other protocols achieving alternative trade-offs between leakage and efficiency have been proposed by e.g. [Cash 2013, Raykova 2012, Pal 2012]. Additional solutions are also possible using techniques for secure multi-party computation [Yao 1982, Ben-Or 1988] but these require further improvement to achieve the required performance.

CMD provides another important technology to this space of solutions (Figure 2) that trade off computational overhead and data leakage and is the first to focus on big data systems.

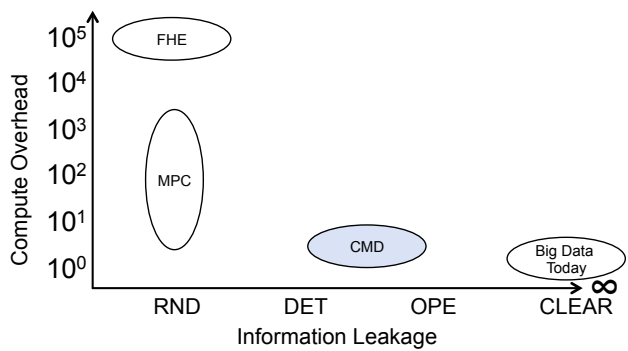

Figure 2. Trade space showing compute overhead versus information leakage for different approaches. RND is a semantically secure encryption scheme (i.e., it leaks no other information other then the length of the message). DET deterministically encrypts each input into exactly one ouput thus leaking equality information. OPE encrypts the data in a way that preserves (and thus reveals) the relative order of the all inputs. CLEAR has no encryption and leaks all information to someone who has access to the data. FHE is fully homomorphic encryption. MPC is multi-party computation. CMD most often uses DET and OPE.

## III. ASSOCIATIVE ARRAYS

CMD operates by working on a mathematical object called an *associative array* that combines features of sparse matrices and triple store databases (such as Apache Accumulo).

Associations between multidimensional entities (tuples) using number/string keys and number/string values can be stored in associative arrays. An implementation of associative arrays can be found in the D4M software package (d4m.mit.edu)[Kepner 2012]. In two dimensions, A D4M associative array entry might be

```
    A('alice ', 'bob ') = 'cited '
or  A('alice ', 'bob ') = 47.0
```

The above tuples have a 1-to-1 correspondence with their triple store representations

```
      ('alice ','bob ','cited ')
or    ('alice ','bob ',47.0)
```

Associative arrays can represent complex relationships in either a sparse matrix or a graph form (see Figure 3). Thus, associative arrays are a natural data structure for performing both matrix and graph algorithms. Such algorithms are the foundation of many complex analytics [Kepner 2011].

Constructing complex, composable query operations can be expressed using simple array indexing of the associative array keys and values, which themselves return associative arrays

| | |
|---|---|
| `A('alice ',:)` | alice row |
| `A('alice bob ',:)` | alice and bob rows |
| `A('al* ',:)` | rows beginning with al |
| `A('alice : bob ',:)` | rows alice to bob |

The composability of associative arrays stems from the ability to define fundamental mathematical operations whose results are also associative arrays. Given two associative arrays A and B, the results of all the following operations will also be associative arrays

```
     A + B    A – B    A & B    A|B    A*B
```

Associative array composability can be further grounded in an axiomatic definition for associative arrays in terms of semi-modules over semi-rings [Kepner 2013b].

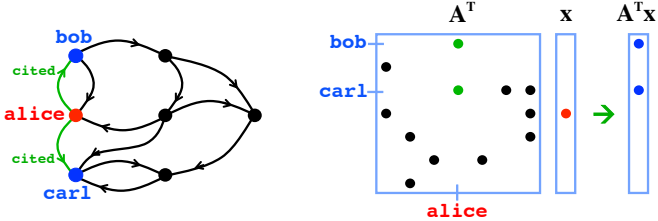

Figure 3. A graph describing the relationship between alice, bob, and carl (left). A sparse associative array **A** captures the same relationships (right). The fundamental operation of graphs is finding neighbors from a vertex (breadth first search). The fundamental operation of linear algebra is vector matrix multiply. D4M associative arrays make these two operations identical. Algorithm developers can use both graph and linear algebra to perform complex operations.

Associative arrays are a critical part of D4M, which has proven useful in a variety of domains (e.g., documents, network logs, social media, and DNA sequences). Measurements using D4M indicate these algorithms can be implemented with a tenfold decrease in coding effort when compared to standard approaches [Kepner 2012]. For example, the D4M schema (aka NuWave) used extensively by the Apache Accumulo community allows diverse data to be quickly ingested and fully indexed using a handful of tables. A feature of the D4M schema is that the data obeys the algebra of associative arrays.

The top part of Figure 4 demonstrates an example of how D4M transforms network log data into an associative array. The raw network traffic logs are collected in a dense tabular form that is made sparse using the same approaches found in the standard D4M schema (i.e., each dense table column and value are appended to make a new sparse table column). CMD exploits this same sparse structure to work on associative arrays that have been encrypted in various forms.

## IV. COMPUTING ON MASKED DATA SYSTEM

The standard CMD use case is as follows. First, users transform their data into associative arrays following the D4M schema (as described in Section III). Then, the components of the associative array's rows, columns, and values are masked using different encryption schemes; this process induces a permutation on rows and columns as they are restructured in lexicographic order by their masks. At this point the masked data structure can optionally be distributed to a system in the encrypted form. Next, algebraic operations are performed on the masked associative arrays. Finally, the results are collected by the user and unmasked.

The bottom part of Figure 4 demonstrates some of the masks that can be used in CMD: DET for the rows (since range queries on rows aren't required), OPE for the columns (which allows for range queries), and RND (a semantically secure encryption scheme) for the values. Another option would be to use an additively homomorphic encryption scheme (HOM+) if the values require summing.

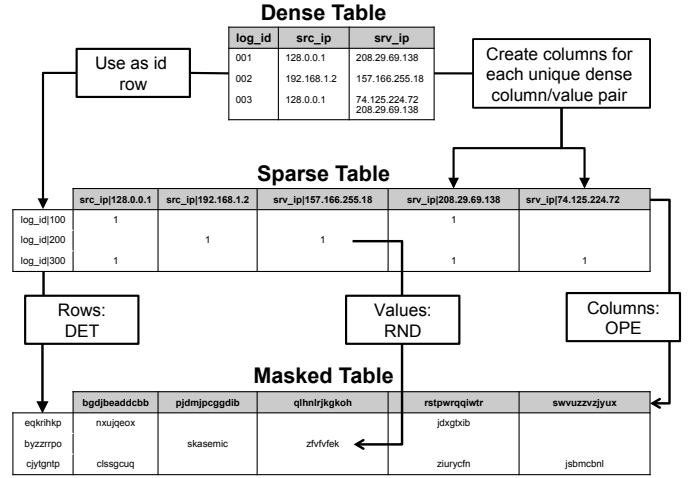

Figure 4. Masking network data records. Dense data is made sparse using the D4M schema that is widely used in the Apache Accumulo community. Dense table column names and values are appended to make columns in the sparse table, which moves most of the semantic content into the rows and columns. The sparse table is then masked using a variety of encyrption schemes depending upon the desired application. The rows are masked using DET, the columns are masked using OPE, and the values are masked using RND.

DET and OPE induce a random permutation on the rows and columns (resp.) of the sparse matrix, but the overall structure of the matrix is preserved. Linear algebra and the algebra of associative arrays are invariant to such permutations (i.e., linear algebraic operations on the masked table will have the same effect as linear algebraic operations on the unmasked sparse table). This is the key concept of CMD, and gives it the ability to perform a wide range of operations on masked data.

Note that CMD may leak the permuted sparse structure of the table. However, determining whether this corresponds to a known unpermuted structure requires solving the graph isomorphism problem, which is not believed to have a generic efficient solution.

Computing the `src_ip` to `srv_ip` graph from the network log data shown in Figure 4 is accomplished by matrix multiplication of different parts of the table. Figure 5 shows the process by which query columns are selected, masked with OPE, and then used to select parts of the table to be multiplied. Associative array algebra allows many possible matrix multiplies to be defined beyond the traditional: +.* (i.e., elements are multiplied and then summed in the usual numerical sense). A particularly useful associative array matrix multiply is the "pedigree preserving" matrix multiply: ∪.& (i.e., elements are first "&" and then unioned). Figure 5 shows how CMD carries out this operation using D4M syntax with the function CatKeyMul performing the pedigree preserving matrix multiply. The result is an associative array depiction of the src_ip to srv_ip graph with the corresponding lists of log_id's stored in the values. In this example, the system holding the table and the system performing the matrix multiply never unmask the data. This is possible since OPE allows the encrypted range query to select the relevant rows for the multiplication and DET encryption of the rows allows for concatenation. The query parameters and the results are masked and unmasked independently from the storage and computation systems.

```
   ┌──── A = CatKeyMul(    T(:,'src_ip|000. : src_ip|999.')).',
   │                       T(:,'srv_ip|000. : srv_ip|999.')) )
┌─────────────────┐
│ Mask column ranges
│    with OPE     │
└─────────────────┘
   │
   └──── A = CatKeyMul(    T(:,'bgdjbeaddcbb : pjdmjpcggdib ')).',
   │                       T(:,'qlhnlrjkgkoh : swvuzzvzjyux ')) )
┌─────────────────┐
│ Pedigree preserving │
│  matrix multiply │
└─────────────────┘
```

| A | qlhnlrjkgkoh | rstpwrqqiwtr | swvuzzvzjyux |
|---|---|---|---|
| bgdjbeaddcbb |  | eqkrihkp cjytgntp | cjytgntp |
| pjdmjpcggdib | cjytgntp |  |  |

Unmask OPE rows/columns and DET values

| A | srv_ip\|157.166.255.18 | srv_ip\|208.29.69.138 | srv_ip\|74.125.224.72 |
|---|---|---|---|
| src_ip\|128.0.0.1 |  | log_id\|100 log_id\|300 | log_id\|300 |
| src_ip\|192.168.1.2 | log_id\|200 |  |  |

Figure 5. Querying and correlating masked data. Query columns are specified by the user and then masked by the user using OPE. Pedigree preserving matrix mulply is then performed on the masked data. The result is unmasked by the user revealing src_ip to srv_ip graph with values corresponding to the log_id.

## V. IMPLEMENTATION & RESULTS

This section describes a prototype system for a subset of CMD and its applications to bioinformatics and social media data. At the moment, the prototype system supports DET encryption using the AES256 [FIPS 197] implementation found in OpenSSL (www.openssl.org) in cipher block chaining mode. The cryptographic key and initialization value[1] used in DET are derived from a user-provided password. To make string handling easier, the results are made into printable characters using Base64 encoding (code.google.com/p/stringencoders).

---

[1]The current implementation of CMD uses the same initialization value for all plaintexts. Unfortunately, this permits equality testing between 16-byte blocks of large encrypted data items. Future versions may use message-dependent initialization values [Popa 2011, Halevi 2003] to provide better security at a small performance cost.

The prototype implementation consists of four new D4M functions to go from plaintext (pt) to masktext (mt)

```
mt  = StrMask(pt,password)      Mask string
pt  = StrUnmask(mt,password)    Unmask string
Amt = Mask(Apt,password)        Mask array
Apt = Unmask(Amt,password)      Unmask array
```

In the current implementation, Mask always performs DET. To mask an array, StrMask is applied independently to each cell in the array. An optional 3rd argument to these functions would allow the CMD system to utilize other encryption primitives like OPE; this will be implemented in future versions.

An example application where masking can be helpful is in the matching of genetic sequence information. This problem is amenable to D4M because matching of genetic sequences can be accomplished with a simple associative array matrix multiply [Kepner 2013c], and it is amenable to CMD because masking can alleviate privacy concerns.

As a demonstration, consider an associative array of genetic DNA data where each row is the name of the sample, each column is unique 10-mer of DNA (i.e., a sequence of 10 bases – a c g t), and each value is the position of that 10-mer in the sequence. For example, a few samples from the National Center for Biotechnology Information (NCBI) GenBank database (www.ncbi.nlm.nih.gov/genbank/) look like:

| Apt | aaaaactaat | aaaaactatt | aaaaacttta | aaaaagaaat | aaaaagaacc |
|---|---|---|---|---|---|
| JN005713.1:AEN70400.1 |  | 597 |  |  |  |
| JN005718.1:AEN70406.1 |  |  |  |  | 300 |
| JN005718.1:AEN70407.1 | 9 |  | 459 |  |  |
| JN033200.1:AEK64751.1 |  | 597 |  |  |  |
| JN851865.1:AFN02593.1 |  |  |  | 332 |  |
| JN851865.1:AFN02596.1 |  |  |  | 176 |  |
| JN851865.1:AFN02601.1 |  |  |  | 38 |  |
| JN851865.1:AFN02604.1 |  |  |  | 309 |  |

Masking all data with DET produces the following:

| Amt | 5ANec Ar+wv 1obhV H3por AA== | B0+EE jaoJT XOYwB K/qXz VQ= | QjPfr MmU+d yoq3y Gzv2r Cw== | RUkkN WI3vL mBdmJ B+5dr Jg== | es5m9 kcE6i BoTMz vdMbe +w== |
|---|---|---|---|---|---|
| 8cYAGF9vJjRMrJz HDbS9ohtbedz41X Du5UbNUxmAM50= |  | fUstngef q1UMYwO8 dQscIg== |  |  |  |
| 8cYAGF9vJjRMrJz HDbS9oiGM/ijoC5 jWGfUnorPzdu0= |  |  |  | QbA5pxfo CbCzuR4m EBIAJg== |  |
| 8cYAGF9vJjRMrJz HDbS9orPYpw0B8S ChdTXTRV6N9nQ= |  |  |  | hYrPmkVx qi9pZzT9 SRgBDw== |  |
| 8cYAGF9vJjRMrJz HDbS9osG32fnb+k 5TbFcKlZ+0G5w= |  |  |  | /ju1MSx2 2AgMr8oP EqFMtQ== |  |
| EPPrWCthJ7q7bfM rgfJo4hq2+TTarA +kVSgfdfbS7KQ= | olhyGImM qfkTh+jv voetlg== |  |  |  |  |
| EPPrWCthJ7q7bfM rgfJo4sNiT++k9B oV4OWP0rIlI60= |  | E/t2jH8X nUhthHun K0zhlQ== |  |  | 4Bb8w6BI KbvJWgqi 6Xg2AQ== |
| GgnerAuBESX1bi/ QbuXApH+WeaWKU6 68H/QpQibHvvY= |  |  | heyVUO8r qBftKX6c 3FVFvA== |  |  |
| tgGLs3km72vPOAK b8VoLG+53S/QZlc Vfa8FKv5y6xfE= |  |  | heyVUO8r qBftKX6c 3FVFvA== |  |  |

Notice that both the rows and columns have been permuted. Finding the matches in this data is accomplished with the following D4M matrix multiply and threshold operation

```
Xmt  = Abs0(Amt) * Abs0(Amt.')    Norm & multiply
Xtop = Xmt > cut                   Threshold
```

where .' denotes the matrix transpose, Abs0 performs the zero-norm on the values (i.e., converts all non-zero values to 1), and `cut` is user defined threshold parameter. The resulting associative array, `Xmt`, holds the masked DNA matches

| Xmt | 8cYAGF9vJjRMrJz HDbS9ohtbedz41X Du5UbNUxmAM50= | 8cYAGF9vJjRMrJz HDbS9oiGM/ijoC5 jWGfUnorPzdu0= | 8cYAGF9vJjRMrJz HDbS9orPYpw0B8S ChdTXTRV6N9nQ= | 8cYAGF9vJjRMrJz HDbS9osG32fnb+k 5TbFcKlZ+0G5w= | EPPrWCthJ7q7bfM rgfJo4hq2+TTarA +kVSgfdfbS7KQ= | EPPrWCthJ7q7bfM rgfJo4sNiT++k9B oV4OWP0rIlI60= | GgnerAuBESX1bi/ QbuXApH+WeaWKU6 68H/QpQibHvvY= | tgGLs3km72vPOAK b8VoLG+53S/QZlc Vfa8FKv5y6xfE= |
|---|---|---|---|---|---|---|---|---|
| 8cYAGF9vJjRMrJz DbS9ohtbedz41XDu 5UbNUxmAM50= | 1 | | | | | | 1 | |
| 8cYAGF9vJjRMrJz DbS9oiGM/ijoC5jW GfUnorPzdu0= | | 1 | 1 | 1 | | | | |
| 8cYAGF9vJjRMrJz DbS9orPYpw0B8SCh dTXTRV6N9nQ= | | 1 | 1 | 1 | | | | |
| 8cYAGF9vJjRMrJz DbS9osG32fnb+k5T bFcKlZ+0G5w= | | 1 | 1 | 1 | | | | |
| EPPrWCthJ7q7bfMr gfJo4hq2+TTarA+k VSgfdfbS7KQ= | | | | | 1 | | | |
| EPPrWCthJ7q7bfMr gfJo4sNiT++k9BoV 4OWP0rIlI60= | 1 | | | | | 2 | | |
| GgnerAuBESX1bi/Q buXApH+WeaWKU668 H/QpQibHvvY= | | | | | | | 1 | 1 |
| tgGLs3km72vPOAKb 8VoLG+53S/QZlcVf a8FKv5y6xfE= | | | | | | | 1 | 1 |

Unmasking the above matrix gives the same result as if the computation had been performed on the plaintext data

| Xpt | JN005713.1:AEN70400.1 | JN005718.1:AEN70406.1 | JN005718.1:AEN70407.1 | JN033200.1:AEK64751.1 | JN851865.1:AFN02593.1 | JN851865.1:AFN02596.1 | JN851865.1:AFN02601.1 | JN851865.1:AFN02604.1 |
|---|---|---|---|---|---|---|---|---|
| JN005713.1:AEN70400.1 | 1 | | | 1 | | | | |
| JN005718.1:AEN70406.1 | | 1 | | | | | | |
| JN005718.1:AEN70407.1 | | | 2 | | | | | 1 |
| JN033200.1:AEK64751.1 | 1 | | | 1 | | | | |
| JN851865.1:AFN02593.1 | | | | | 1 | 1 | 1 | |
| JN851865.1:AFN02596.1 | | | | | 1 | 1 | 1 | |
| JN851865.1:AFN02601.1 | | | | | 1 | 1 | 1 | |
| JN851865.1:AFN02604.1 | | | 1 | | | | | 1 |

Figure 6 shows the performance of the various parts of the computation as a function of the size of the DNA data. The runtime of the masked computation is within a factor of 2 of the runtime of the plain computation across the entire range. The increase in execution time is due to the increase in the data size that comes from padding the data to 16 bytes blocks for encryption and Base64 encoding. This doubles the size of the data and results in a doubling of computation time. For many big data systems, doubling the computation time in exchange for taking the data out of the clear is a reasonable tradeoff.

The masking time at small data sizes is dominated by the overhead of our implementation that currently uses file IO. Future implementations can use direct procedure calls to remove this overhead. At larger data sizes, the masking time approaches a constant overhead of 0.6 times the compute time. The majority of this mask time arises from the permutation of the resulting associative array to maintain lexicographic order. In applications where this is not required, this step can be eliminated. Thresholding of the values makes the output DNA match matrix (`Xtop`) much smaller than the input DNA matrix and so the unmask time is small and is dominated by the file IO of the implementation. As with the masked computation time, future implementations can use direct procedure calls to remove this overhead.

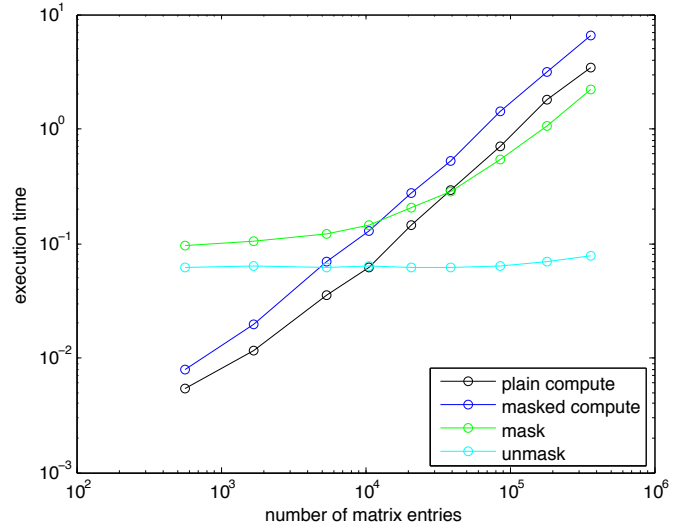

Figure 6. Performance of CMD on finding matches in DNA data. The graphs shows the execution time (in seconds) versus the number on non-zero entries in the matrix, which corresponsds to the number DNA sequences in the data. The masked compute time is always within a factor of 2 of the plain compute, which is the performance goal of the CMD approach. The masking and unmasking time are relatively small compared to the compute time.

The CMD system has also been implemented to work with the D4M package and Apache Accumulo database. D4M operations to insert and query associative arrays to and from an Accumulo database remain unchanged. In order to insert masked data, a user would issue a mask command to the data prior to inserting. To query for masked data, a user would send a masked query and unmask the resultant associative array. Consider a user who wishes to use the CMD system to store and retrieve Twitter data stored in an Accumulo database.

To insert an associative array `A` into an Accumulo table `T` with password `password` using the CMD system, the user would first mask `A` using `password` and then insert into `T` using the D4M put command. For example:

```
Amt = Mask(A,password)       Mask A
put(T,Amt)                   Insert Amt into T
```

After inserting, table `T` now contains the masked elements of `A`. To query the table `T` for keyword `word`, the query must be done with the masked version of `word` using the same `password` during insert. Further, the query will return an associative array whose elements are masked. In order to view the plaintext, this associative array will need to be unmasked using the same `password` as before. For example:

```
Amt=T(:,StrMask(word,password))   Query
Apt=Unmask(Amt,password)          Unmask
```

A prototype Twitter corpus was used to test the performance of the CMD on database operations such as inserting and querying. The performance of using the CMD system vs. standard plaintext operations is summarized in Figure 7. The left y-axis corresponds to insertion times and the right y-axis corresponds to query times.

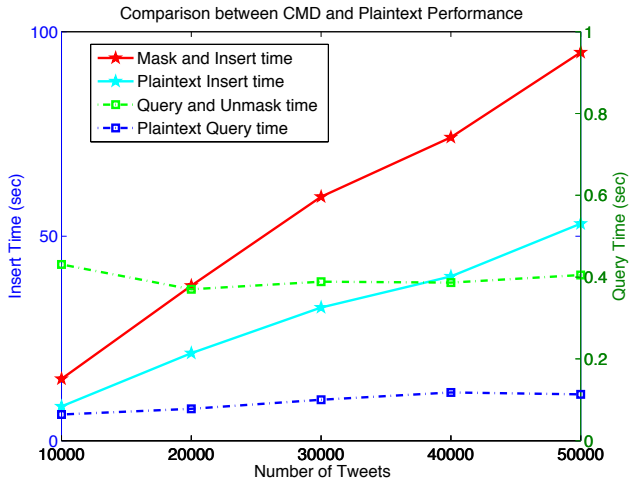

Figure 7. Performance of CMD system on a varying number of Tweets. The Left Y-axis corresponds to the time taken for Inserting data. The red line corresponds to the CMD insert operation which includes time for masking and inserting. The cyan line corresponds to just inserting plaintext. The red and cyan lines show the relative performance between CMD and plaintext and corresponds to an overhead less than 2x. The right Y-axis corresponds to the time taken for querying for data. As querying in Accumulo is a constant time operation, both lines are relatively flat. The difference between the green and blue dashed lines correspond to the overhead incurred in masking the query and unmasking the result associative array.

Figure 7 shows the perforomance obtained by the CMD system for a varying data size from 10,000 tweets to 50,000 tweets. The Twitter data inserted is represented by the D4M schema. The insert times include time taken for inserting (and masking) the data stored in all the tables required by the D4M Schema. These operations are dominated by the database insert and query time. The increase in insert and query time is due to the increase in the data size that comes from padding the data to 16 bytes blocks for encryption and base64 encoding. This doubles the size of the data and results in a doubling of insert and query time. This is a reasonable tradeoff for many big data systems in exchange for no longer having data in the clear.

## VI. SUMMARY & FUTURE WORK

CMD is a novel approach that increases big data veracity while allowing a wide range of computations and queries to be performed with low overhead. The encryption schemes used in CMD can be tailored to the application: semantically secure, deterministic, order preserving, and homomorphic encryption can be used in complementary fashion to provide the best overall application solution. Through continued dialog between big data application designers and cryptographers, further improvements can fine-tune the balance between performance and veracity of big data systems.

A prototype CMD was implemented with DET by adding four functions to the existing D4M library. Using this implementation, CMD was demonstrated on a DNA sequence matching problem and the performance met the 2x performance goal. The CMD system was also demonstrated with a social media dataset that also met the performance goals. Future improvements to the prototype include expanding and improving the encryption schemes to use the full set developed in CryptDB and using direct procedure calls to reduce the overhead of the mask and unmask functions.